\documentclass[12pt]{article}
\usepackage{epsf}

\newlength{\dinwidth}
\newlength{\dinmargin}
\def\lapproxeq{\lower .7ex\hbox{$\;\stackrel{\textstyle <}{\sim}\;$}}
\def\gapproxeq{\lower .7ex\hbox{$\;\stackrel{\textstyle >}{\sim}\;$}}

\def\be{\begin{equation}}
\def\ee{\end{equation}}
\def\bea{\begin{eqnarray}}
\def\eea{\end{eqnarray}}

\def\gtrsim{ \;\raisebox{-.7ex}{$\stackrel{\textstyle >}{\sim}$}\; }
\def\lesim{ \;\raisebox{-.7ex}{$\stackrel{\textstyle <}{\sim}$}\; }

\begin{document}
\titlepage

\begin{flushright}
IPPP/04/26 \\
DCPT/04/52\\
2nd July 2004 \\
\end{flushright}

\vspace*{4cm}

\begin{center}
{\Large \bf The spread of the gluon $k_t$--distribution and the determination of the
saturation scale at hadron colliders in resummed NLL BFKL}

\vspace*{1cm} \textsc{V.A.~Khoze$^{a,b}$, A.D. Martin$^a$, M.G. Ryskin$^{a,b}$ and W.J. Stirling$^{a,c}$} \\

\vspace*{0.5cm} $^a$ Department of Physics and Institute for
Particle Physics Phenomenology, \\
University of Durham, DH1 3LE, UK \\[0.5ex]
$^b$ Petersburg Nuclear Physics Institute, Gatchina,
St.~Petersburg, 188300, Russia \\[0.5ex]
$^c$ Department of Mathematical Sciences, 
University of Durham, DH1 3LE, UK \\%
\end{center}

\vspace*{1cm}

\begin{abstract}
The transverse momentum distribution of soft hadrons and jets that accompany central hard-scattering
production  at hadron colliders is of great importance, since it has a direct bearing on the ability
to separate new physics signals from Standard Model backgrounds.
 We compare the predictions for the gluonic $k_t$--distribution using two different approaches:
resummed NLL BFKL
 and DGLAP evolution. We find that as long as the initial and final virtualities ($k_t$) 
along the emission chain are not too close to each other,  the NLL resummed BFKL results do 
not differ significantly from those obtained using standard DGLAP evolution.
The saturation momentum $Q_s(x)$, calculated within the resummed
BFKL approach, grows with $1/x$ even slower than in the leading-order DGLAP
case.
\end{abstract}

\newpage

\section{Introduction}
The high-energy behaviour of QCD amplitudes is described by the
BFKL/CCFM equation~\cite{BFKL,CCFM} which sums to all orders
leading logarithmic contributions of the form 
$(\alpha_s\;\ln s)^n$.
 The next-to-leading logarithmic (NLL) corrections
 $\alpha_s(\alpha_s\; \ln s)^n$ have been calculated in Refs.~\cite{NLL}. The NLL
corrections turn out to have a large numerical effect, and therefore in order to obtain
reliable predictions one needs to first understand the origin of the large
numerical coefficients at NLL and then if possible to further resum the main part of 
the NLL contribution.

In the past few years there have been a number of studies of the BFKL approach at NLL accuracy, see
Refs.~\cite{AV,Cif,BKL}, which concentrate on the
properties and the behaviour of the gluon Green function.
In this paper we focus instead on the gluon transverse momentum
 ($k_t$) distribution along the BFKL evolution chain and
compare the predictions obtained using both the resummed NLL BFKL and the
LO DGLAP evolution approaches.

Besides the pure theoretical interest in further understanding the properties of QCD in
the high-energy limit, there are also important phenomenological implications.
Precise knowledge of the $k_t$ distribution 
of intermediate (i.e. accompanying) gluons is important for achieving a  
better understanding of the structure of the so-called `underlying event' at
hadron colliders such as the Tevatron and the LHC. As has been emphasized
 in \cite{Stalk}, the fact that the transverse momentum of the emitted 
partons (and therefore of the soft hadrons and minijets) grows with energy 
(i.e. with $1/x$) could cause problems due to a `noisy' underlying event affecting the
extraction of a clean new-physics signal (for example, inclusive production of 
Higgs bosons, SUSY particles, etc.) at the LHC. 
Therefore it is very important to be able calculate the predicted gluon $k_t$ distribution
as precisely as possible. Of course the standard approach is to use 
parton shower Monte Carlos such as HERWIG and PYTHIA to estimate the $k_t$ distribution
of accompanying partons/hadrons. However as these models are based on DGLAP evolution, they
 do {\it not} contain all the expected
logarithmic contributions in the high-energy limit. It is therefore important to understand
 how and why the DGLAP and (potentially more realistic) 
resummed NLL BFKL approaches differ. Such a comparison will form
a major part of our study. Also very relevant, is the
high-energy behaviour of the
saturation momentum $Q_s$ --- the transverse momentum at which
non-linear effects (from gluon-gluon rescattering and recombination)
become important and start to saturate the parton densities.

The BFKL amplitude is usually described in terms of the Pomeron intercept
 (i.e. the singularity in the complex momentum $j$--plane)
$j=1+\omega(\gamma)$ which depends on the anomalous dimension
($\gamma$) of the eigenfunction. In contrast,  DGLAP evolution
is usually described in terms of $\gamma(\omega)$, i.e. the anomalous
dimension is considered as a function of $\omega$, the conjugate variable to $x$.

 In order to compare the DGLAP and BFKL predictions, in Section~2 we will consider the
unintegrated gluon distribution $f_g(x,k_t,\mu)$  --- that is, the probability to find a
gluon carrying longitudinal momentum fraction $x$ and transverse
momentum $k_t$ in a process with  (hard) factorization scale $\mu$ --- written 
in the form of a double contour integral over both
$\gamma$ and  $\omega$, i.e.  performing simultaneous Mellin transforms with respect
to both the $k_t$ and $x$ distributions. Depending on which of these integrations is performed first, 
one obtains either the standard DGLAP or the BFKL form.

In Section~3 we introduce a resummation which is a modification of that proposed by the Firenze
group~\cite{Cif2,SPol}. The idea is to modify the contibution of the
poles at $\gamma=0$ ($\gamma=1$) corresponding to the normal (inverse)
$k_t$--ordered DGLAP evolution. To be explicit, we include the full LO
DGLAP (splitting function) contribution, allow for the appropriate
`energy' variable $x=Q^2/s$, and account
for energy-momentum conservation.   After these modifications have been made,
it has been shown that the remaining part of the NLL correction
does not exceed 7\% of the original value.

Solving the BFKL equation for the intercept
\begin{equation}
j=1+\omega(\gamma)=1+ {\alpha_s N_c\over \pi }\; \chi(\gamma)
\end{equation}
 with respect to
 $\gamma$, one obtains the NLO contribution $C_n(\alpha_s/\omega)^{n+1}$
 to the DGLAP anomalous dimension $\gamma(\omega)$, equivalently the 
$C_n x^{-1} \alpha_s^n \ln^{n-1}(x)$ contribution to the (gluon-gluon) splitting function.
 It turns out that in leading-order (LO)
BFKL the first two non-trivial terms vanish\footnote{Note that Eq.~(\ref{eq:X0}) below can be written in the form
$X_0=1/\gamma+2\Psi(0)-\Psi(1+\gamma)-\Psi(1-\gamma)=1/\gamma + O(\gamma^2)$.},
i.e. $C_1=C_2=0$~\cite{C12}.
Therefore the expected BFKL corrections to DGLAP evolution are numerically 
rather weak (see \cite{AF} for a more detail discussion). On the other
hand, in terms of the intercept $j$ the value of $\omega(\gamma)$ given by
the LO BFKL equation is much larger than that coming from DGLAP. At
first sight this looks like a contradiction. However the situation
changes dramatically after the NLL resummation. Now the BFKL intercept becomes
close to, and even slightly smaller than, the corresponding DGLAP quantity. 
Thus in any kinematic
configuration where the initial and final virtualities (transverse
momenta) are not too close to each other, the DGLAP and the BFKL
predictions do not differ significantly (in agreement with the
conclusions of Ref.~\cite{AF}).

The dependence of the  $k_t$ distribution on the overall event kinematics is discussed in 
Section~4 and the $x$-dependence of the saturation momentum $Q_s(x)$ is discussed in Section~5.
Section~6 contains our conclusions.

 \section{The unintegrated gluon distribution}
\label{sec:unint}
We begin by considering the unintegrated (over transverse momentum) gluon distribution 
$f_g(x,k_t,\mu)$,  in the form \cite{UNINT}
\begin{equation}
f_g(x,k_t,\mu)\; =\; T_g(k_t,\mu)\; \frac{\alpha_s(k_t^2)}{2\pi}\; \int_x^{1-\Delta}dz\; P_{gg}(z)
\;(x/z) g(x/z, k_t^2)
\end{equation}
where the survival probability $T$ is given by
\begin{equation}
T_g(k_t,\mu) = \exp\left( -\int_{k_t^2}^{\mu^2} \frac{d{k'}_t^2}{{k'}_t^2}  \; 
\frac{\alpha_s({k'}_t^2)}{2\pi}\; \int_0^1 dz' \;  \Big(\Theta(z'-\Delta)\Theta(1-z'-\Delta)z'P_{gg}(z')+n_F P_{qg}(z')\Big) \right).
\label{eq:T}
\end{equation}
Here the infrared cutoff is $\Delta = k_t/(\mu + k_t)$;
$\mu$ is the factorisation scale; and $n_F$ is the number of active quark flavours.
When $k_t$ is close to $\mu$ we may
neglect the double logarithms in the $T$-factor, since
$\alpha_s \ln^2(k_t/\mu) \ll 1$, and thus $T\simeq 1$.
Then the unintegrated distribution is simply related to the derivative of the standard 
(integrated) gluon
parton distribution function:
\begin{equation}
f_g(x,k_t) = \left[ {d\over d \/ \ln\mu^2}\; xg(x,\mu^2)\right]_{\mu = k_t}.
\label{eq:derivative}
\end{equation}
Introducing the Mellin-transform variables, $\gamma$ and $\omega$,
conjugate to $k^2_t$ and $1/x$ respectively, we can write
$f_g$ as
\begin{equation}
f_g(x,k_t)=\int^{+i\infty}_{-i\infty} \frac{d\gamma}{2\pi i}
   \int^{+i\infty}_{-i\infty} \frac{d\omega}{2\pi i}
F(\gamma,\omega)\frac{e^{\omega Y+\gamma r}}{\omega-\bar a
X(\gamma,\omega)}
\label{eq:doubleint}
 \end{equation}
where $\bar a = N_c\alpha_s/\pi$, $Y=\ln(1/x)$ and $r=\ln(k_t^2/Q^2_0)$.
$F(\gamma,\omega)$ represents the input distribution at $k_t^2=Q^2_0$, where $Q_0$ is the starting
scale for perturbative evolution assumed to be $\sim 1$~GeV.
$X$ is the resummed BFKL intercept, $X=X_0+\bar a X_1$, where the leading order (LL)  contribution is
\begin{equation}
X_0(\gamma)=2\Psi(1)-\Psi(\gamma)-\Psi(1-\gamma)
\label{eq:X0}
\end{equation}
 and $X_1$ is the NLL correction.
The contour integration over $\omega$ goes to the right of all singularities,
while the real part of the anomalous dimension $\gamma$ is bounded by  $0 < {\rm Re}\gamma <1$.

In the DGLAP limit $r \gg Y$ we close the $\gamma$ contour around the pole
given by  $1/(\omega-\bar aX)$, leaving the inverse Mellin integral over $\omega$
with $\gamma=\gamma(\omega)$. In the BFKL case ($Y \gg r$) we close the
$\omega$ contour around the same pole $1/(\omega-\bar aX)$ and write the
result as an integral over $\gamma$, with  $\omega=\omega(\gamma)$.
In the latter form
\begin{equation}
f_g=\int^{+i\infty}_{-i\infty} \frac{d\gamma}{2\pi i}e^{\gamma r +
\omega_s Y}
 B(\gamma)
\label{eq:dgamma}
  \end{equation}
with $\omega_s(\gamma)$ given by the solution of the equation
\begin{equation}
 \omega-\bar aX(\gamma,\omega)=0
\label{eq:solution}
\end{equation}
and 
\begin{equation}
 B(\gamma)={F(\gamma,\omega_s(\gamma)) \over 1-X'_\omega} ,  \qquad
X'_\omega= {\partial X(\gamma,\omega) \over \partial\omega}\Big|_{\omega = \omega_s(\gamma)}  .
\end{equation}
 With this representation, the input distribution at $Q_0$ is absorbed into $B(\gamma)$,
 which may therefore be fitted to reproduce the data.

Another possibility is to use the conventional Mellin ($\omega$)
representation for $f_g$ by closing the $\gamma$ contour around the nearest pole.
Strictly speaking the
 BFKL function $X_0$ of Eq.~(\ref{eq:X0}) has poles at each integer $\gamma$.
The pole at $\gamma=0$ corresponds to the normal twist-2 DGLAP contribution,
the pole at $\gamma=1$ corresponds to inverse $k_t$ ordering
($k_t \ll Q_0$).  The other poles at $\gamma=-1,\, -2,\, ...(\gamma=2,\, 3,\, ...)$ are the
higher-twist contributions (3, 4, ... gluons in the $t$-channel), corresponding to
normal (inverse) $k_t$ ordering, hidden in the
Reggeization of the BFKL gluons. However in practice we know that the higher-twist
contribution at small $x$ is small (see, for example, Ref.~\cite{MRSTerror2}).
Therefore, to begin, we may neglect the poles at negative $\gamma$, arguing
that phenomenologically the input function
$F(\gamma,\omega)$ is numerically small at negative 
integer values of $\gamma$.\footnote{Analogously, in the $d\gamma$ integral (\ref{eq:dgamma}) we cannot rule out
the possibility of other singularities in the $\omega$ plane situated to the left of the
leading pole at $\omega = \omega_s$.  However, since in any case we cannot justify the BFKL
approach for large $|\omega| \sim 1$, in the present context we only keep the leading pole in
(\ref{eq:dgamma}).}
We then obtain
\begin{equation} f_g=\int^{+i\infty}_{-i\infty} \frac{d\omega}{2\pi i},
e^{\gamma_s(\omega) r+\omega Y}D(\omega) 
\label{eq:domega}\end{equation}
 where
\begin{equation} D(\omega)=\gamma_s(\omega) M(\omega)
\label{eq:Domega}
\end{equation}
and
$M(\omega)=\int_0^1 x^{\omega} g(x,Q^2_0)dx$ is the known Mellin transform
($x$-moments) of the input gluon distribution. The first factor $\gamma$ in 
(\ref{eq:Domega}) is due to the derivative in (\ref{eq:derivative}), and all $\gamma$ dependent quantities are evaluated at  $\gamma=\gamma_s(\omega)$, the solution of
(\ref{eq:solution}).

\section{The resummed NLL BFKL intercept}
In this section we consider the resummed NLL contribution to the unintegrated gluon distribution defined in the previous section.
We use the idea proposed in Ref.~\cite{SPol}, but with a small modification,
which is rather simple and more convenient for our purpose.
The crucial point is the fact that the major part of the {\cal O}($\alpha_s$)
 correction is actually contained in the {\cal O}($\omega$) 
contribution\footnote{Recall that in the BFKL approach, $\omega \sim \alpha_s$.}.
Next we note that the nearest, and most important, poles in the quantity $X$ of
Eq.~(\ref{eq:X0}) at
$\gamma=0$ and $\gamma=1$ correspond, respectively, to the well known, normal- ($k_t \gg Q_0$)
and inverted- ($k_t \ll Q_0$) ordered, twist-2 DGLAP
contributions. Thus the leading-order characteristic function $X_0$ contains
two parts.  One is the twist-2 poles $1/\gamma$ and $1/(1-\gamma)$, and the
remainder is the higher-twist component, $X_0^{\rm (ht)}$.  That is
\begin{equation}
X_0~=~\frac 1{\gamma}+\frac 1{1-\gamma}+X_0^{\rm (ht)},
\label{eq:X0parts}
\end{equation}
where, from (\ref{eq:X0}), we have
\begin{equation}
X_0^{\rm (ht)}~=~2\Psi(1)-\Psi(1+\gamma)-\Psi(2-\gamma).
\label{eq:ht}
\end{equation}

First, we have to modify the poles
to include in the residues the {\it full} LO DGLAP splitting
function. Because in the physical (axial) gauge both
the BFKL and DGLAP LO contributions are given by ladder-type
Feynman diagrams, the Mellin transform of the 
 final (modified) amplitude in the $\omega$, $\gamma$ representation may
be written in the same exponential form as 
Eqs.~(\ref{eq:doubleint},\ref{eq:dgamma},\ref{eq:domega}).
Thus the residue 1 in the twist-2 DGLAP pole at $\gamma=0$ is replaced
by the full DGLAP splitting function
\begin{equation} \frac
1{2N_c}\omega P_{gg}(\omega)=1+\omega A_1(\omega)\, .
\end{equation}
In pure gluodynamics (i.e. $n_F=0$) $A_1=-\frac{11}{12}
+O(\omega)$. To account for the quark loop contribution
for $n_F\neq 0$ we have to replace $A_1$ by
\begin{equation}
A_1(\omega)\; +\; n_F\left( {\bar a \over 4N^2_c\gamma}\; P_{gq}(\omega)
P_{qg}(\omega)\; -\; {1 \over 3}\right),
\end{equation}
 where the first term in the brackets corresponds to the
real quark two-loop contribution\footnote{The $k_t$ integral over the second loop,
in the DGLAP approximation, generates the $1/\gamma$ pole.}, as indicated by the extra $\alpha_s$
factor. The second term is the virtual quark
 loop insertion in the gluon propagator, and corresponds to the $P_{qg}$ term in
 the $T$ factor of Eq.~(\ref{eq:T}). Formally, the NLL correction is represented by the
 leading ($\omega$) term in $A_1$. Nevertheless here we keep the full
 $\omega$ dependence of the LO DGLAP splitting kernels.
 The same procedure can be followed to modify the residue of the $1/(1-\gamma)$ pole,
 which corresponds to the DGLAP evolution with inverse $k_t$ ordering.

Another modification is necessary.
In order to compare the BFKL and DGLAP  predictions we have to write everything in terms of
the DGLAP `energy' variable $x=Q^2/s$. Due to the
asymmetry in the definition of the energy scales, for the normal and inverse
ordered contributions, we need to correct the
contribution of the $1/(1-\gamma)$ pole, and to
replace $(1-\gamma)$
by $(1-\gamma+\omega)$ (see Section~3.3 and Eq.~(60) of Ref.~\cite{SPol}) in the
terms that are singular when $\gamma$ is close to 1.

Putting everything together, we finally obtain the characteristic function
\begin{equation}
X(\gamma,\omega)=X_0+\bar aX_1,
\label{eq:Xequals}
\end{equation}
where the LL intercept $X_0$ is given by (\ref{eq:X0}).
The NLL contribution
\begin{equation}
\bar aX_1=\left(\frac{1+\omega A_1(\omega)}{\gamma}-\frac 1{\gamma}
+ \frac{1+\omega A_1(\omega)}{1-\gamma+\omega}-\frac
1{1-\gamma}\right)-\omega X_0^{\rm (ht)},
\label{eq:X1equals}
\end{equation}
consists of a twist-2 part, which is shown in brackets,
and a correction, $-\omega X_0^{\rm (ht)}$, to the higher-twist component (\ref{eq:ht}) of $X_0$,
whose origin we now explain.
Following Ref.~\cite{EKL}, we have multiplied the higher-twist contribution (\ref{eq:ht}) by
a factor $(1-\omega)$, which effectively accounts for the kinematical
constraints and provides conservation of energy and momentum.  At $\omega =1$
the whole contribution vanishes.  Note that the twist-2 part already satisfies this condition,
since we use the full DGLAP splitting function.

\begin{figure}
\begin{center}
\centerline{\epsfxsize=\textwidth\epsfbox{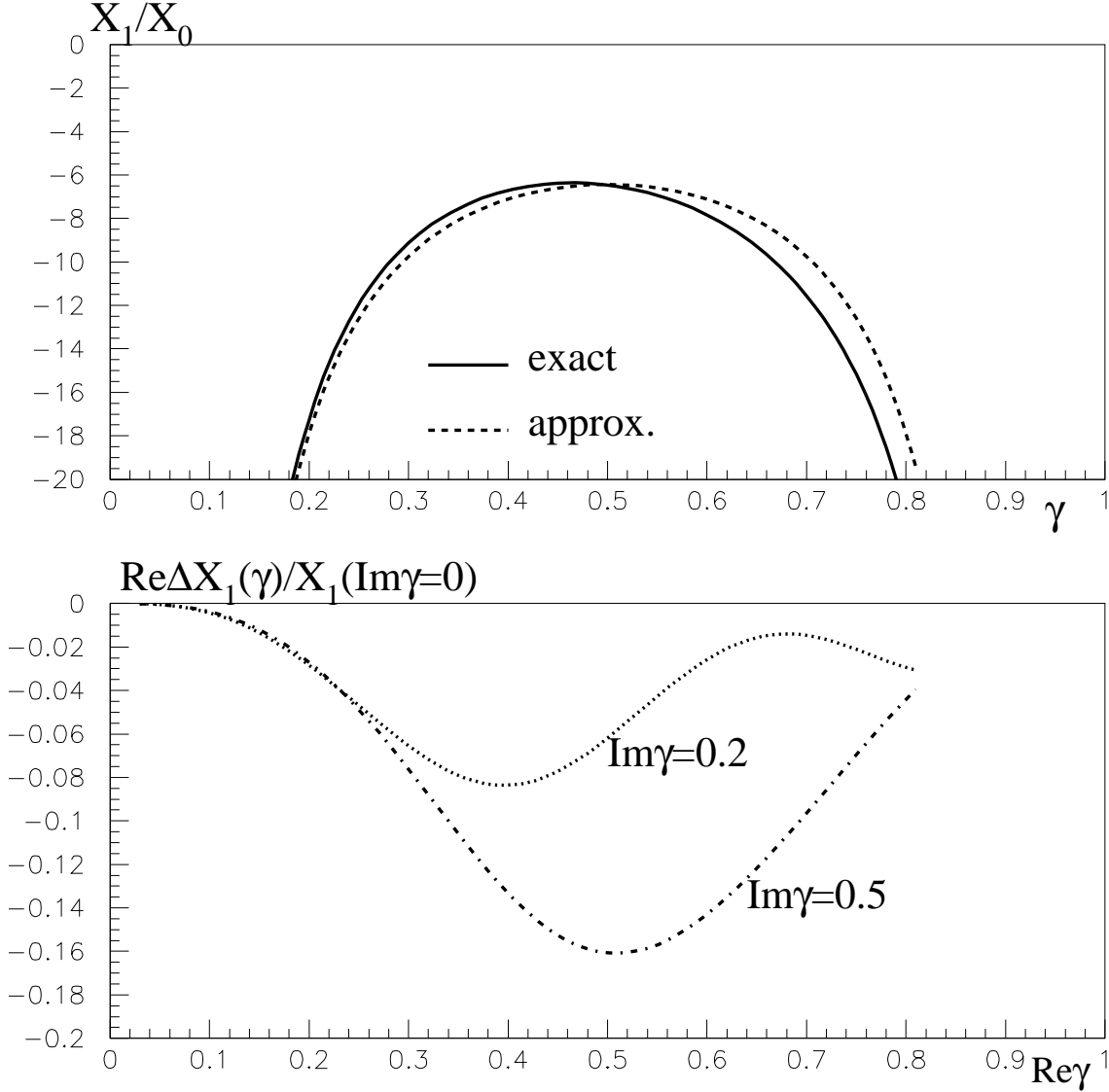}}
\caption{The comparison of the approximation of Eqs.~(\ref{eq:ht}-\ref{eq:X1equals})
for the characteristic function with the exact NLL result.  The upper diagram shows $X_1/X_0$
as a function of $\gamma$ with Im $\gamma =0$, whereas lower diagram shows the deviation, $\Delta X_1$,
suitably normalised, of our approximation of $X_1$
from the exact result for various values
of Im $\gamma$.   Recall that $\omega = \bar a X_0 + \bar a^2 X_1$, where $\bar a =3\alpha_s/\pi$.}
\label{fig:check}
\end{center}
\end{figure}

We have checked that, in the important region $0<\gamma<0.6$,
the approximation given by Eqs.~(\ref{eq:ht}-\ref{eq:X1equals})
reproduces the known exact NLL BFKL {\cal O}($\alpha_s$) result
to within $7\%$ accuracy.  This is illustrated in Fig.~\ref{fig:check}.
For real $\gamma$, the continuous curve in the upper plot is the result of the
exact calculation of $X_1/X_0$ for $n_F=0$ (Eqs. (53,54) of Ref.~\cite{SPol}, see also
Refs.~\cite{NLL}), whereas the dashed line comes from the approximation
given by Eqs.~(\ref{eq:ht}-\ref{eq:X1equals}) above.  We need to know
the intercept not only at the saddle point (Re $\gamma \sim 0.3-0.5$, depending on the
kinematics), but also in a region in the complex plane around this point (with $|{\rm Im} \gamma| \lesim 0.5$).
We therefore show in the lower plot of Fig.~\ref{fig:check} the deviation, $\Delta X_1$, of our approximation of $X_1$
from the exact result\cite{NLL} for various values of Im $\gamma$, normalized to the
value of $X_1$ for $\gamma={\rm Re}~\gamma$ with Im $\gamma$ = 0.

Moreover, note that the approximation of
 Eqs.~(\ref{eq:ht}-\ref{eq:X1equals}) can be used even for
rather large values of $\alpha_s$, since from Eq.~(\ref{eq:solution}) we always have $\omega_s > 0$.
Note that by using the above form we are effectively expanding in $\omega$, which here
plays the role of a small parameter, rather than in the coupling $\alpha_s$ (i.e. $\bar
a$).\footnote{With the exception of the quark loop contribution,
see footnote 4.}  This is the result of resumming the
NLL ({\cal O}($\alpha_s$)) corrections to the leading-order BFKL/CCFM
intercept.

\begin{figure}
\begin{center}
\centerline{\epsfxsize=\textwidth\epsfbox{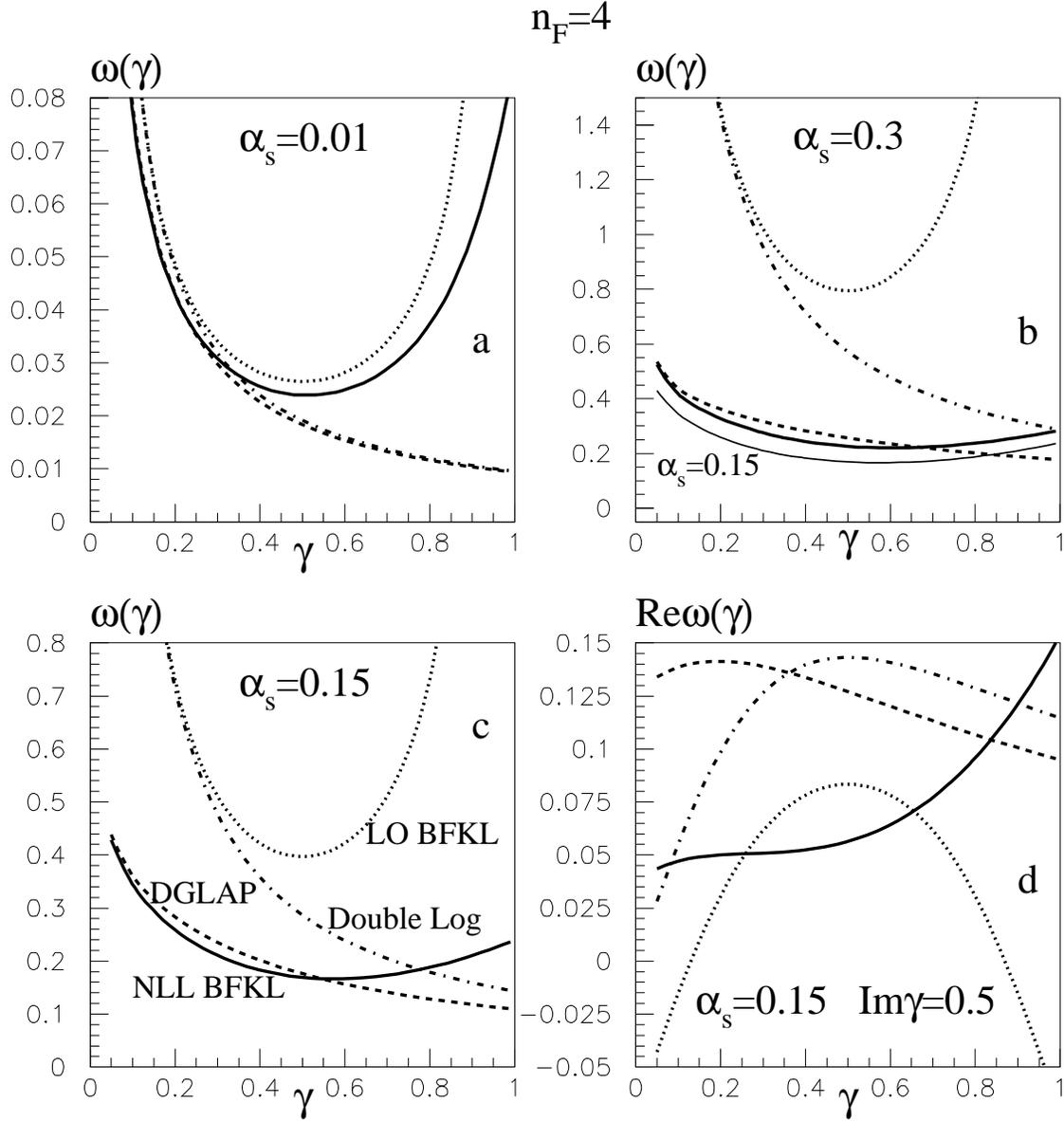}}
\caption{The `Pomeron' intercept, $\omega_s$, obtained by solving Eq.~(\ref{eq:solution}),
in the following four approximations: LO BFKL; DL; NLL BFKL and LO DGLAP; as described in
the text.  We include quark-loop corrections, with $n_F=4$.}
\label{fig:w1}
\end{center}
\end{figure}

\begin{figure}
\begin{center}
\centerline{\epsfxsize=\textwidth\epsfbox{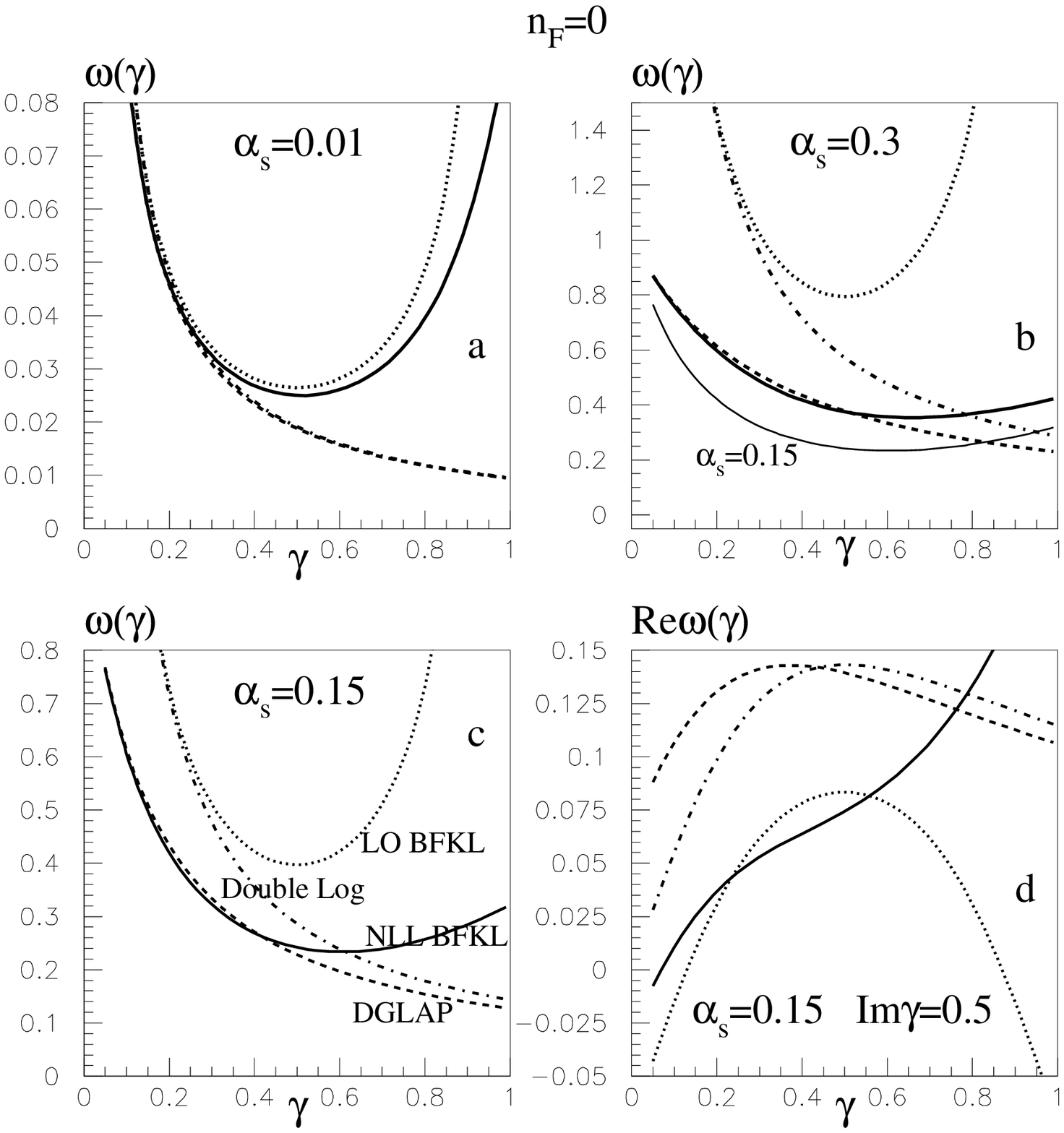}}
\caption{As for Fig.~\ref{fig:w1}, but for pure gluodynamics, with $n_F=0$.}
\label{fig:w2}
\end{center}
\end{figure}

The solutions of Eq.~(\ref{eq:solution}) for the leading singularity
$\omega_s(\gamma)$ are shown in Figs.~\ref{fig:w1} and \ref{fig:w2}.
We compare the results obtained using four different approximations:
\begin{itemize}
\item[(i)] the dotted line corresponds to
the well-known LO BFKL function $X(\gamma,\omega)=X_0(\gamma,0)$;
\item[(ii)] the Double Logarithmic (DL) contribution $X=1/\gamma$, which is of course the same
for the LO DGLAP and LO BFKL cases, is shown by the dot-dashed line;
\item[(iii)] the solution of (\ref{eq:solution}) with the full resummed function
$X(\gamma,\omega)$ given by Eqs.~(\ref{eq:ht}-\ref{eq:X1equals}) is shown by the solid line;
\item[(iv)]
the dashed line shows the pure DGLAP result where we keep just the
same LO DGLAP contribution that was included in (\ref{eq:X1equals}), that is
$X=(1+\omega A_1(\omega))/\gamma$.
\end{itemize}
At very small $\alpha_s=0.01$ the solid (resummed NLL BFKL) curve is
rather close to the dotted one (LO BFKL). However already at
$\alpha_s=0.15$ the NLL corrections significantly change the behaviour of
$\omega_s$ for real $\gamma$ (Figs.~\ref{fig:w1}a,b,c and Figs.~\ref{fig:w2}a,b,c). In particular, the
solid (NLL BFKL) curve becomes closer to the dashed (DGLAP) curve. Note that
the resummed value of $\omega_s$ depends weakly on the QCD coupling
$\alpha_s$. The NLL BFKL solutions for $\alpha_s=0.3 (0.15)$ shown in Figs.~\ref{fig:w1}b and \ref{fig:w2}b
by the heavy (thin) solid lines are quite similar.

>From the plots, we see that in the important region, Re $\gamma < 0.6$, the resummed NLL BFKL
intercept is  very similar to that for the DGLAP case.  In fact for
$\gamma \sim 0.3$ the NLL BFKL value is even a little below the DGLAP intercept.  However for
large values of $\gamma \gtrsim 0.6$, the NLL BFKL curves go above the DGLAP curves,
due an additional positive term from the pole
 $1/(1-\gamma+\omega)$ which arises from the inverse $k_t$-ordered contribution.
The approximate equality of the NLL BFKL and DGLAP curves for $\gamma < 0.6$ occurs
because this positive contribution is compensated by the virtual corrections
corresponding to gluon Reggeization, that is by the higher-twist poles
at $\gamma=-1, -2, ...$ and $\gamma=2+\omega,3+\omega, ...$, which give a
negative contribution.

As we go into the complex $\gamma$ plane, the resummed intercept decreases faster than that
of DGLAP, and is closer to the original LO BFKL, since we become further away from the
twist-2 DGLAP pole at $\gamma=0$.   This decrease of the NLL intercept improves the
convergency of the saddle-point integral.
To demonstrate this behaviour of $\omega_s$ in
 the complex $\gamma$ plane, we plot in Figs.~\ref{fig:w1}d and \ref{fig:w2}d the results for
 Im $\gamma=0.5$. The value of the intercept decreases with increasing Im $\gamma$
 and, due to the higher twist contributions, we indeed see that in the BFKL case it
decreases faster than in the DGLAP case.

 Finally, we can investigate the role of the quark loop corrections.
 In Fig.~\ref{fig:w1} we have considered the (realistic) case of four light quarks
 ($n_F=4$). In Fig.~\ref{fig:w2} we show the analogous results for pure
 gluodynamics ($n_F=0$). The inclusion of quark loop contributions evidently
 shifts the position of the singularity at $\omega_s$ to the left; the absolute value of the (negative)
 virtual quark loop correction is larger than the real quark loop
 contribution.  This demonstrates that the gluon spends part of the `evolution time'
 (i.e. rapidity interval) in the form of a quark-antiquark state which
 corresponds to a lower intercept $j < 1$, that is to a negative
value of $\omega_s$.

\section{The evolution of the gluon $k_t$ distribution}

In this section we use the formalism developed above to study numerically
the gluon $k_t$ distribution, and in particular how it changes as one moves
along the evolution chain starting from a given input distribution at a particular value of $x$. We begin
by introducing the Green function ($G$) corresponding to the
unintegrated gluon distribution with the initial condition
$\delta(1-k^2_t/k'^2)$ at $x=x'$, i.e. a fixed non-zero value of $k_t$ at $x=x'$. According to the results of the previous section, we have
\begin{equation}
G(r,r';x,x')=\int_{-i\infty}^{+i\infty}\frac{d\gamma}{2\pi i}
e^{\gamma(r-r')+\omega_s(Y-y')}.
\label{eq:Green}
\end{equation}
Here $r'=\ln(k'^2/Q^2_0)$, $y'=\ln(1/x')$ and $\omega_s=\omega_s(\gamma)$
is the solution of (\ref{eq:solution}).

\begin{figure}
\begin{center}
\centerline{\epsfxsize=\textwidth\epsfbox{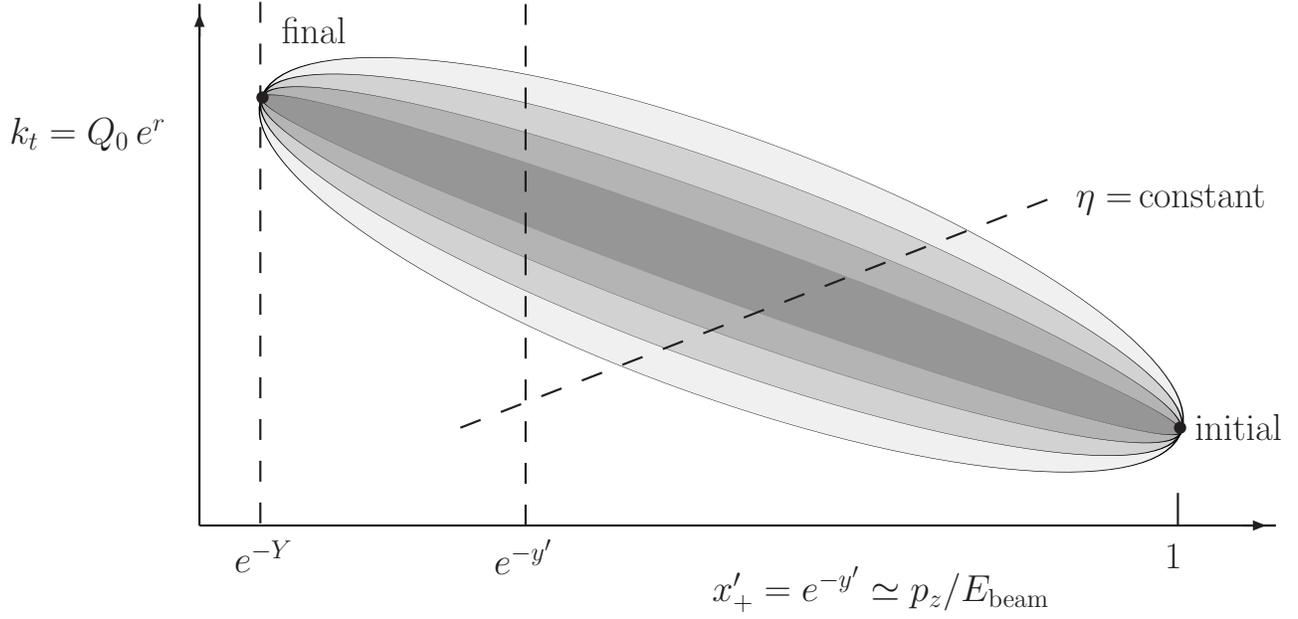}}
\caption{A schematic sketch of the `cigar-like' kinematic domain in $x'$ and $k_t$
covered by the possible paths from the initial
to the final evolution points. Note that the scales for $k_t$ and $x'_+$ are both logarithmic,
as indicated by the exponential factors.  Also note that the + subscript on $x'$
indicates a light cone variable, which is only introduced to ensure
that the fixed $\eta$ curve is a straight line. Figs.~\ref{fig:fixed} and \ref{fig:running}a,b
show the gluon $k_t$ distribution at fixed values of $y'$, whereas Fig.~\ref{fig:running}c,d show
the distributions at fixed $\eta$.}
\label{fig:cigar}
\end{center}
\end{figure}

To evaluate the spread in the  transverse momentum $k'_t$ at some intermediate
value of $y' < Y$  (that is $x' > x$), we write the final unintegrated distribution
$f_g(x,k_t)$ as the convolution
\begin{equation}
f_g(x,k_t) =\int \frac{d^2k'_t}{\pi k'^2_t} f_g(x',k'_t)
 G(r,r';x,x')
\label{eq:convolG}
\end{equation}
and study the distribution of the integrand over $k'_t$.
It is simplest to elucidate the procedure pictorially.
The complete evolution is shown schematically in Fig.~\ref{fig:cigar}.  We have the source at $x=1$
and $k_t=Q_0$, and consider the evolution to the final gluon at $x={\rm exp}(-Y)$
and some large value of $k_t$, say 30 GeV.  This point can be reached by different
evolution trajectories, indicated by the shaded area in the diagram.
Of course the exact form of the shaded region will depend on the
approximation used to calculate the behaviour of the gluon along the
evolution chain.  For example, we shall see that, as expected, LO BFKL
with running coupling $\alpha_s$, tends to populate
the low $k_t$ domain much more than the other three approximations that we use.

We are now in a position to study the correlated quantity of experimental interest,
that is the distribution of intermediate gluons in a process where a high $p_t$ (or high
mass) particle is produced, with (light-cone) momentum fraction
$x_+=x={\rm exp}(-Y)$.  The scale of the process specifies the final $k_t$ of
our gluon evolution.   We have therefore
written, in Eq.~(\ref{eq:convolG}), the final gluon distribution as a convolution in $k'_t$ of the gluon density
at $x'$ and the Green function (\ref{eq:Green}) which describes the evolution from $x'$
to the final point $x$. The quantity of experimental interest is thus the integrand,
 $f_g(x',k'_t)G(r,r';x,x')$, which represents the $k'_t$ distribution at an intermediate value of $x'$.
 In particular we wish to investigate
how the intermediate $k'_t$ distributions, obtained in the various approaches,
differ from one another; or, to be more explicit, to see if the NLL BFKL result
brings the LO BFKL behaviour close to the DGLAP result.
It is convenient to plot the results as the
ratio of the integrand to the final gluon density, $f_g(x',k'_t)G(r,r';x,x')/f_g(x,k_t)$.
This corresponds to the `normalized' distribution $ k'^2_t\, dN/d k'^2_t$ of the intermediate
gluons as a function of  $\ln k'^2_t$, at the chosen intermediate value $x'={\rm exp}(-y')$.

For the purpose of illustration we calculate the evolution of the gluon $k_t$ distribution taking
 the initial scale to be $k_t=Q_0=1.2$~GeV and choosing the
final $k_t=30$~GeV at $Y=\ln (1/x)=8$.
For the LHC energy ($\sqrt s=14$~TeV) this corresponds to
a $E_T=30$~GeV jet with pseudorapidity $\eta \sim -2$ (in the
direction opposite to the parent proton), which can be observed in the
central detector. If we set $y'=4$, we can study the
$k'_t$ distribution of the accompanying jets with longitudinal momentum
$p_z=E_{\rm beam}\, e^{-y'}\sim 55$~GeV (in the parent proton direction).

\begin{figure}
\begin{center}
\centerline{\epsfxsize=\textwidth\epsfbox{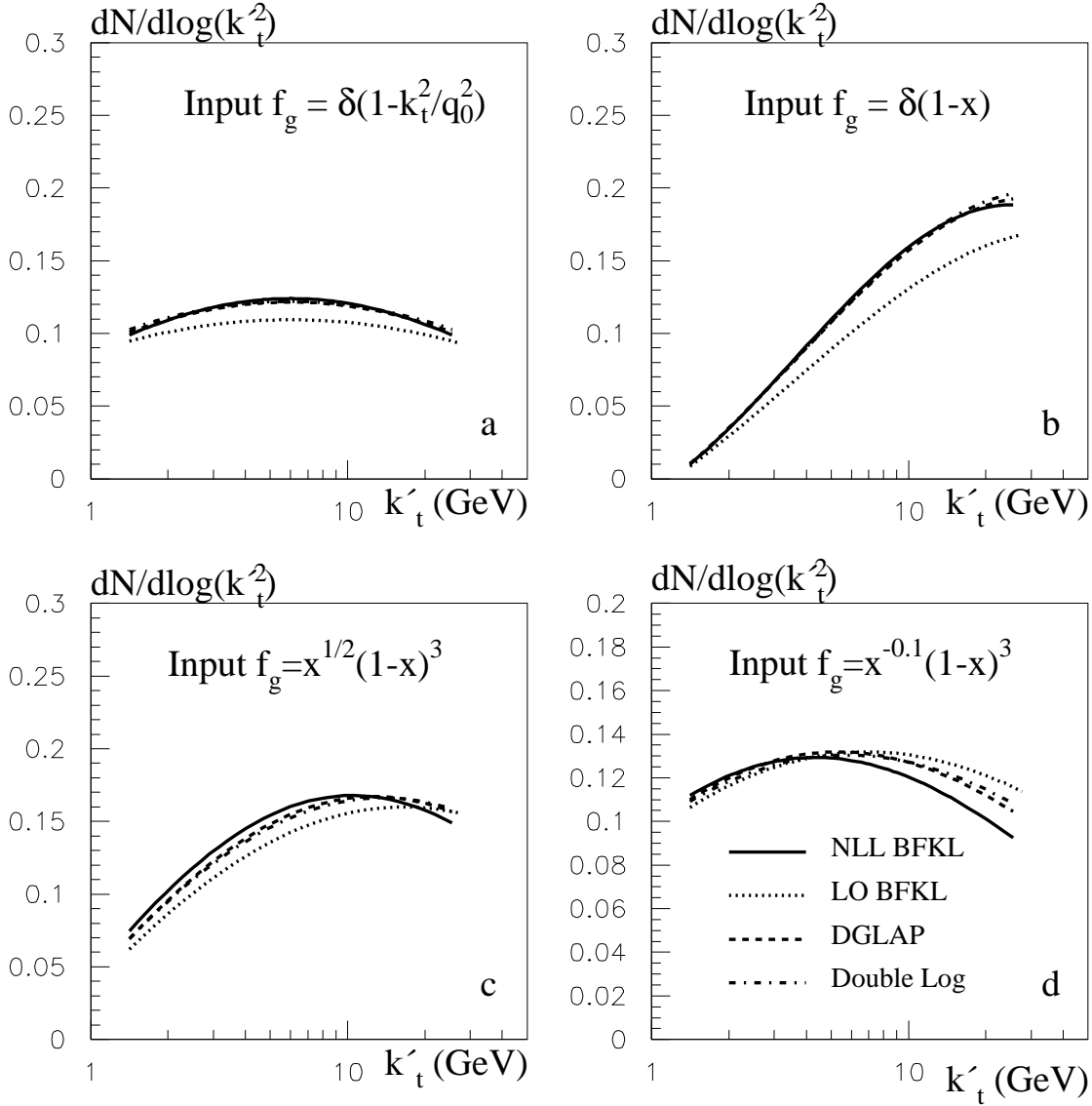}}
\caption{The $ k'^2_t\, dN/d k'^2_t$ distributions of the intermediate
gluons as a function of  $\ln k'^2_t$ at  intermediate $y'=4$,
for fixed $\alpha_s=0.2$.  We show the four distributions corresponding to
using LO BFKL, resummed NLL BFKL, DGLAP and DL, labelled just as in
Figs.~\ref{fig:w1} and \ref{fig:w2}.  The four plots show the distributions
for different choices of the input distributions.}
\label {fig:fixed}
\end{center}
\end{figure}

The shape of the distribution depends on the initial and boundary
conditions. Starting with $f_g=\delta(1-k^2_t/q^2_0)$ at $x=1$
 (that is from the conditions corresponding to the BFKL Green function)
we obtain almost flat distributions (see Fig.~\ref{fig:fixed}a) in all four cases: LO
BFKL, resummed NLL BFKL, DGLAP and DL. Note that we use the same labelling for the
curves as in Figs.~\ref{fig:w1} and \ref{fig:w2}.  Using instead the input $f_g=\delta(1-x)$ at $k_t=Q_0$
we obtain distributions that grow almost linearly with $\ln k_t$, see Fig.~\ref{fig:fixed}b.
 This is caused by the boundary condition
$f_g(x,k_t=Q_0)=0$ for any $x<1$. Again the distributions in all four
cases are close to each other.

Next, in Figs.~\ref{fig:fixed}c,d we show the results for more realistic input distributions:
\begin{equation}
\mbox{valence-like}\;\; f_g(x,k_t=Q_0)=\sqrt x (1-x)^3\;\;\; \mbox{
(Fig.~\ref{fig:fixed}c), }
\label{eq:vallike}
\end{equation}
 and
\begin{equation}
\mbox{Pomeron-like}\;\; f_g(x,k_t=Q_0)=x^{-0.1}(1-x)^3\;\;\;\mbox{
(Fig.~\ref{fig:fixed}d).}
\label{eq:pomlike}
\end{equation}
All the curves in Fig.~\ref{fig:fixed} are calculated using a fixed $\alpha_s=0.2$ and
$n_F=0$.

For the valence-like input (\ref{eq:vallike}), the initial gluons are
concentrated at rather large $x\sim 1$ and the results at low $k'_t$ look
qualitatively similar to those in Fig.~\ref{fig:fixed}b. For the Pomeron-like input
(\ref{eq:pomlike}) we
have a bigger contribution from the small-$x$ and low-scale $k_t$ region.
Therefore the distributions are more close to those in Fig.~\ref{fig:fixed}a.

To fix the input condition  at $k_t=q_0$, the usual procedure for DGLAP evolution, it
is easier to work with the $\omega$ representation, replacing  the contour
integration in the $\gamma$ plane (\ref{eq:dgamma}) by the integral in the $\omega$ plane
(\ref{eq:domega}). We have checked that  to within a few per cent accuracy 
both representations give the same function $f_g$ even for the case when we keep
only the leading pole in the denominator of the integrand (\ref{eq:doubleint}).
Strictly speaking this can only be true for $n_F=0$. For a non-zero $n_F$
there is a second pole in the $\gamma$ plane which corresponds to the second
eigenfunction in the singlet DGLAP evolution where we have two
equations --- one for each of the gluon and quark distributions. In Section~3
 we considered the gluon distribution only, and this second (quark)
equation reveals itself as a new pole at a $\gamma$ value comparable with that
($\gamma_s(\omega)$) of the leading pole.
To obtain a  precise solution we therefore have to keep the contribution from this second pole as well.
However in order to simplify the discussion (and computations)
in this section, we will consider the case of $n_F=0$
only.

\begin{figure}
\begin{center}
\centerline{\epsfxsize=\textwidth\epsfbox{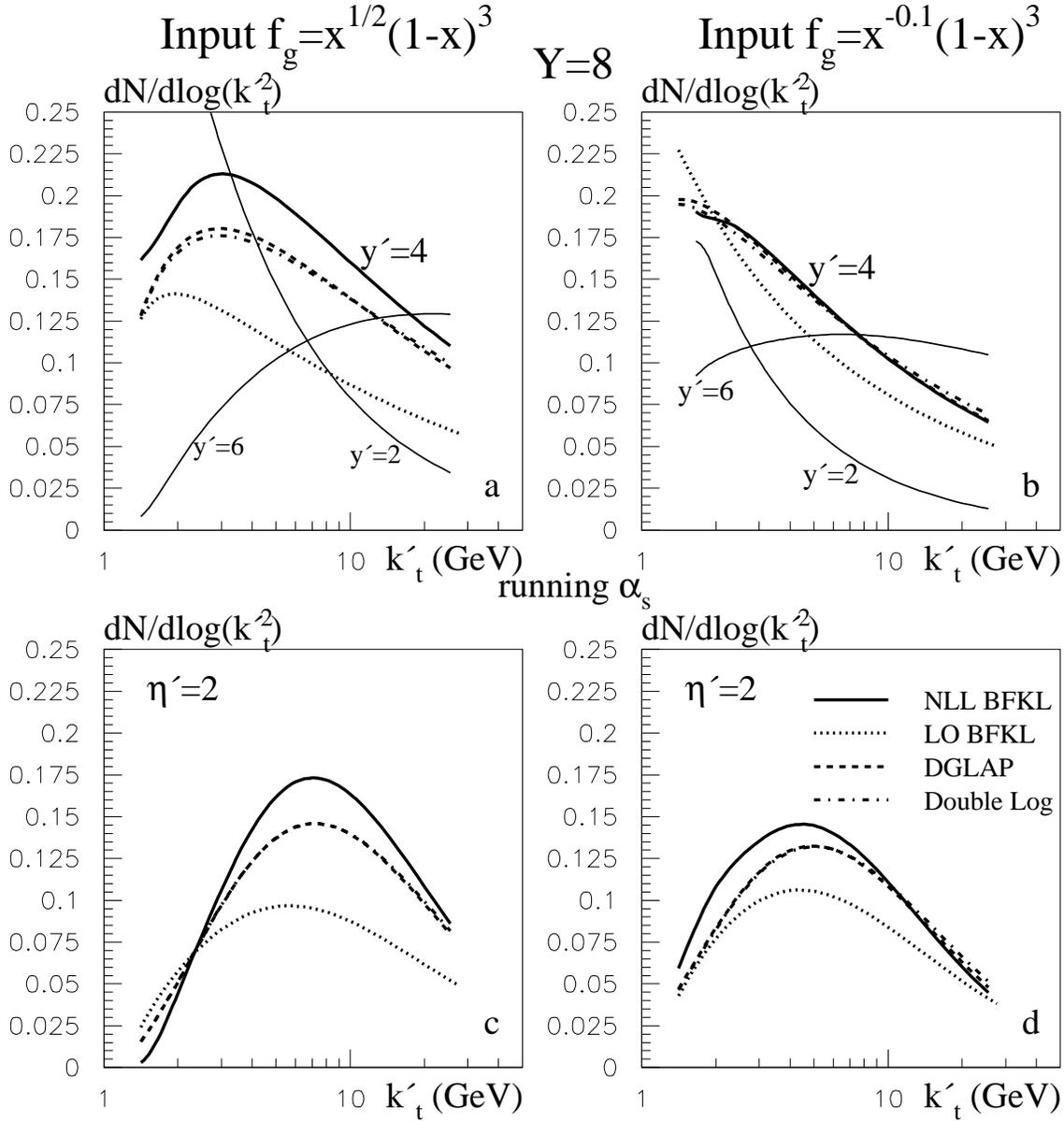}}
\caption{Plots a,b are as for Fig.~\ref{fig:fixed}c,d but with running $\alpha_s$ --
also shown on these two plots (by thin lines) are the NLL BFKL distributions for the intermediate values $y'=2$
and 6, as well as the $y'=4$ distribution which is shown by the
thick line.  Plots c,d show the gluon distributions for a fixed
value of pseudorapidity, $\eta'=2$, rather than fixed $y'$.}
\label{fig:running}
\end{center}
\end{figure}

One advantage of using the representation where we integrate over $d\omega$ is
that, as in the DGLAP case,
we can include the running of $\alpha_s(k'^2_T)$ simply by
replacing the product $\gamma_s(\omega)r$ in the power of the exponent in (\ref{eq:Green}) and (\ref{eq:domega})
 by the integral
\begin{equation}
\gamma_s(r-r')\;\rightarrow\; \int_{r'}^r \gamma_s(\omega,r'')dr''
\end{equation}
where $r'$ is the logarithm of the initial virtuality and
the anomalous dimension $\gamma_s(\omega,r'')$ is calculated using the
running QCD coupling $\alpha_s(r'')$.  Note that in (\ref{eq:domega}) $r'=0$, of course.
The results for running $\alpha_s$ are presented in Fig.~\ref{fig:running} for
 valence-like (\ref{eq:vallike}) (Figs.~\ref{fig:running}a,c) and Pomeron-like (\ref{eq:pomlike})
(Figs.~\ref{fig:running}b,d) inputs.
In addition to the distributions at $y'=4$ (that is, at fixed
longitudinal momentum $p_z\sim 55$~GeV) which are shown in Figs.~\ref{fig:running}a,b,
we also show, in Figs.~\ref{fig:running}c,d,
the distributions at fixed gluon jet (pseudo)rapidity
$\eta'=2$ in the parent proton direction.

As expected, including the running $\alpha_s$ shifts the distributions to lower $k_t$
values where the coupling is larger. This is most evident for the LO BFKL
curve in Fig.~\ref{fig:running}b, where the (Pomeron-like) input does not suppress the
low $k_t$  region. After the resummation of the NLL corrections,
the results depend less on the value of $\alpha_s$, and
the  $\ln k^2_t$ distribution of the intermediate gluons becomes broader, with
the maximum shifted to a larger $k_t$ in comparison with the
LO BFKL distribution.  This is a result of the weaker dependence of the resummed
intercept on the value of $\alpha_s$, see Figs.~\ref{fig:w1} and \ref{fig:w2}.
On the other hand, we see that the form of the NLL $k'_t$ distribution is
very similar to that for DGLAP, since the intercepts $\omega(\gamma)$ are
almost equal in the region of the saddle points.

In Figs.~\ref{fig:running}c,d we present the $k_t$ distributions of jets
with a fixed  pseudorapidity $\eta'=2$.
In this case a small-$k_t$ jet has a smaller longitudinal momentum
(that is, a larger $y' = \ln(1/x')$). As $k_t$ increases we therefore move from 
a distribution with a larger $y'$ to a distribution corresponding to a smaller $y'$. 
To illustrate this, we look at the fixed $y'$ plots, Figs.~\ref{fig:running}a,b, with $y'=4$.
On these plots we also show (by thin lines) the NLL BFKL distributions at
two other fixed values, $y'=2$ and $y'=6$. It is clear that the distributions at fixed $\eta'$ have a maximum,
since the curves with a larger $y'$ grow with $k_t$, while for a smaller $y'$
they decrease. In other words,
the low $k_t$ domain
is suppressed since at fixed $\eta$ the gluon with a smaller $k_t$  has
a smaller longitudinal momentum $p_z$.
 In this case the low $k_t$ part of the distribution corresponds to an extreme
evolution trajectory, where first the major part of the
$\ln(1/x)=Y$ interval is used for BFKL evolution and followed then by an increase
due to DGLAP evolution at almost constant $x$; see Fig.~\ref{fig:cigar}.  Of
course such evolution trajectories give a much smaller overall contribution than those
where both $1/x$ and $k_t$ grow simultaneously leading to double
logarithmic $(\alpha_s\ln(1/x)\ln k_t)^n$ contributions.

The effect of inverse $k_t$-ordering is more evident at fixed $\eta'$.
For the Pomeron-like input, the NLL distribution is broader than the DGLAP
one (see Fig.~\ref{fig:running}d).  However, for valence-like input, it
is a bit narrower (see Fig.~\ref{fig:running}c).  The reason is as follows.
The cigar-like shape of Fig.~\ref{fig:cigar} is broader for BFKL, so we
would anticipate a broader $k'_t$ distribution.   However at fixed $\eta'$,
for low $k_t$ we enter the low $x'$ region where the valence-like input
vanishes.  This has a larger effect on the more spreadout BFKL distribution
than on that of DGLAP, with its thinner cigar which practically does not
sample the region with $k'_t \sim Q_0$.
Evidently, however, the effect is not very strong and
the final NLL BFKL and LO DGLAP distributions do not differ significantly.

This result is of great phenomenological importance, since it implies that for the kinematic regions described above,
which are typical of what may be studied at the LHC, parton shower
Monte Carlos based on LO DGLAP evolution should give a reasonable
approximation to the predictions obtained using a full NLL resummed BFKL calculation.

\begin{figure}
\begin{center}
\centerline{\epsfxsize=\textwidth\epsfbox{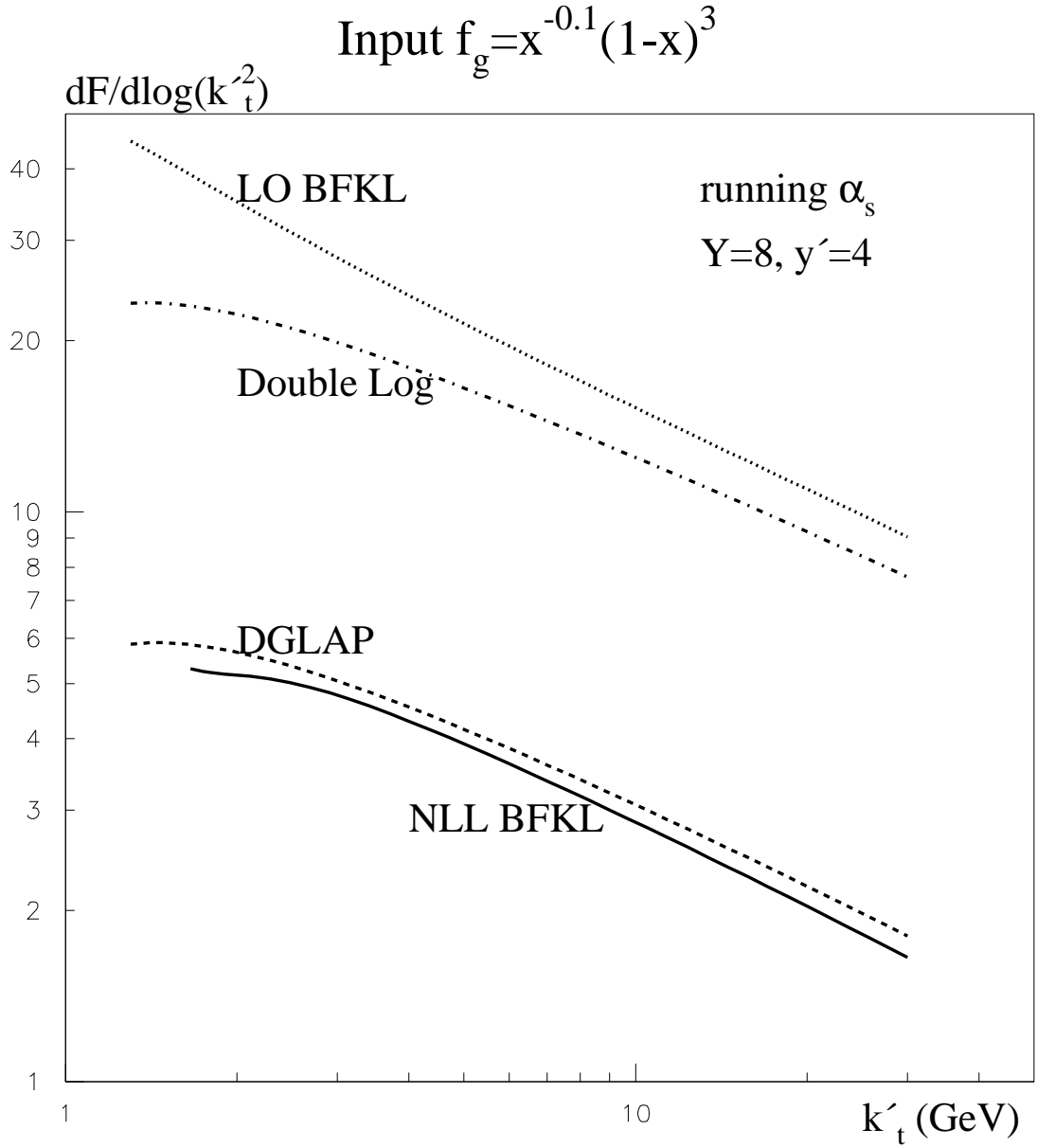}}
\caption{The gluon distributions corresponding to Fig.~\ref{fig:running}b.  Here the intermediate
gluon distribution has not been normalized to the probability to find the final
gluon with transverse momentum $k_t$ and $x={\rm exp}(-Y)$.}
\label {fig:actual}
\end{center}
\end{figure}

Recall that the $k'_t$ distributions have been shown in
Figs.~\ref{fig:fixed} and \ref{fig:running} in `normalized' form,
\begin{equation}
\frac{dN(x',k'_t)}{d{\rm ln}k_t^{'2}}~=~\frac{f_g(x',k'_t)G(r,r';x,x')}{f_g(x,k_t)}.
\end{equation}
That is we have shown the density of intermediate gluons with momentum fraction
$x'={\rm exp}(-y')$ in events where the final gluon has been detected with momentum $k_t$
and fraction $x$.  Therefore we have introduced the factor $f_g(x,k_t)$ in the denominator.
To obtain more insight, we show, in
Fig.~\ref{fig:actual}, the actual $k'_t$ distributions
\begin{equation}
\frac{dF(x',k'_t)}{d{\rm ln}k_t^{'2}}~=~f_g(x',k'_t)G(r,r';x,x')
\label{eq:F}
\end{equation}
obtained in the four
approximations, corresponding to Fig.~\ref{fig:running}b.  This quantity, (\ref{eq:F}),
represents the double inclusive cross section -- the probability to observe both the final
gluon $(x,k_t)$ and the intermediate gluon $(x',k'_t)$.

\section{Determination of the saturation momentum, $Q_s$}

Note that up to now we have been dealing with  single ladder graphs and
therefore have not taken any account of absorptive corrections (or gluon recombination).
These effects are formally suppressed by a factor $1/R^2Q^2$, where $R >
1/Q_0$ is some dimensionful parameter coming from the confinement domain.

However it is well known that at very small $x$ the parton densities grow and the non-linear
two- or multi-ladder contributions must eventually become important. This occurs when
the cross section for gluon-gluon recombination becomes comparable with
the proton area $\pi R^2$, i.e. $\sigma\sim \pi R^2$. Now at high energies the cross 
section behaves as $\sigma\sim  x^{-\omega}/Q^{2(1-\gamma)}$.
Therefore to have $\sigma\sim \pi R^2$ requires
\begin{equation}
\label{eq:s1}
Q^2~=~Q^2_s~\propto ~x^{-\omega/(1-\gamma)}\ .
\end{equation}
This raises the question as to what value of $\gamma$ (and correspondingly $\omega(\gamma)$)
should one use in (\ref{eq:s1})?

 It can be shown that
multi-ladder (fan) graphs becomes important for  $\gamma >
\gamma_c$, where $\gamma_c$ is given by the equation
\begin{equation}
-\frac{d\omega_s(\gamma_c)}{d\gamma_c} =
\frac{\omega_s(\gamma_c)}{1-\gamma_c} \; .
\label{eq:sat}
\end{equation}
This equation was first obtained in Ref.~\cite{GLR}
using the `wave front' method
in the framework of the GLR equation. Then it was considered in Ref.~\cite{bl}.  The equation
can also be justified on the basis of the Balitsky-Kovchegov equation (in conformal
symmetric form) \cite{BVR,RP}. In the Appendix we show how (\ref{eq:sat}) can be derived using BFKL eigenfunctions.

In order to solve Eq.~(\ref{eq:s1}) for $\gamma_c$, we return to the $\gamma$-integral representation of the
amplitude given in Eq.~(\ref{eq:dgamma}).
By evaluating the integral using the saddle point method, we see that
 for large $r$ (i.e. a large interval of $\ln k_t^2$ DGLAP evolution)
the value of the saddle point, $\gamma_{\rm sp} \sim \sqrt{\alpha_sY/r}$, is small, but
increases with the rapidity interval $Y$.
When the position of the saddle point is such that $\gamma_{\rm sp} <\gamma_c$, the
non-linear contributions are negligible. However if $\gamma_{\rm sp} >
\gamma_c$, then the absorptive effects become large and we rapidly reach the
saturation regime with $f_g(x,k_t)\propto k^2_t$.

The value of $\gamma_c=0.37$ was obtained in \cite{GLR} using the
LO BFKL formalism.  Evaluating $\gamma_c$ using the
resummed NLL BFKL expressions 
for $\omega_s$ given in Section~2,  we arrive at 
almost the same value --- $\gamma_c=0.32$
 for $\alpha_s=0.3$ --- and, in fact, in the limit $\alpha_s\to 0$ the
 critical anomalous dimension $\gamma_c\to 0.37$.

\begin{figure}
\begin{center}
\centerline{\epsfxsize=\textwidth\epsfbox{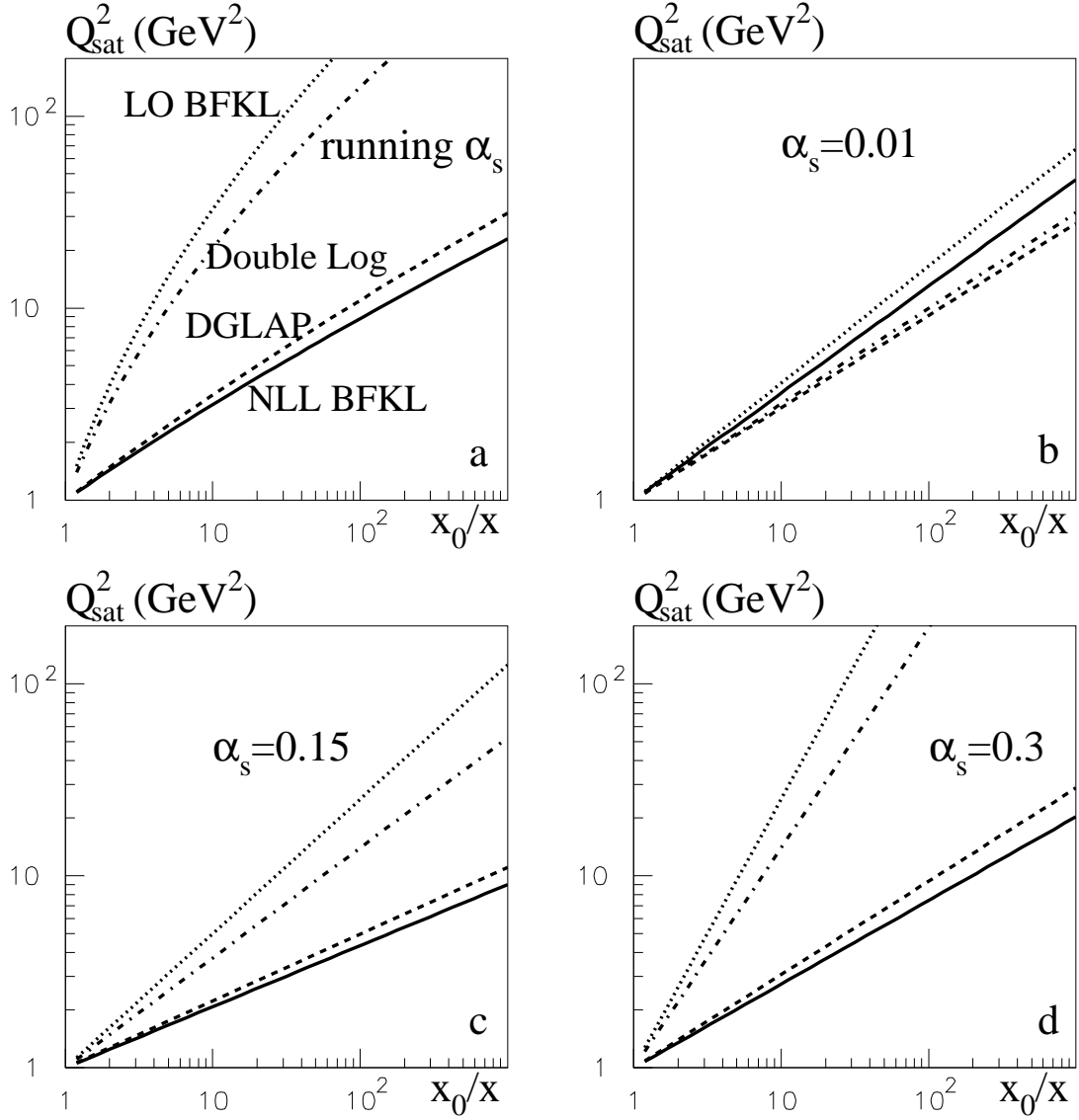}}
\caption{The saturation momentum, $Q_s$, versus $1/x$. Note that the
$1/x$ scale is proportional to $x_0$, which, by definition, is the value
of $x$ at which $Q_s=1$ GeV.   We discuss in the text that an order
of magnitude estimate of $x_0$ is $10^{-4}$.}
\label{fig:sat}
\end{center}
\end{figure}

It is straightforward to obtain the $x$ dependence of the saturation momentum
$Q_s(x)$.  Let us assume that we already reach the value of $Q_s=1$~GeV 
at some $x_0$. With decreasing $x$ we need to increase $Q_s$
 in order to keep the saddle point at $\gamma_{\rm sp}=\gamma_c$.
 The equation which gives the
 saddle point of the integration in Eq.~(\ref{eq:dgamma}) is
\begin{equation}
r=\ln Q^2_s=-Y\frac{d\omega_s}{d\gamma}=Y\frac{\omega_s}{1-\gamma},
\label{eq:speq}
\end{equation}
where we have used (\ref{eq:sat}) to get the last equality.
We cannot determine the saddle point directly from Eq.~(\ref{eq:speq}), since the value of $\omega_s$
depends on the running of $\alpha_s$.  Therefore we need to obtain a
differential equation.
 This leads to
\begin{equation}
\frac{d\ln Q^2_s}{d\ln (1/x)}=\frac{\omega_s(\gamma_c)}{1-\gamma_c}.
\end{equation}
Solving this equation for $\gamma_c$ leads to the $x$ dependence of the
saturation momentum $Q_s$ that is shown in Fig.~\ref{fig:sat}.
Here we include the light quark contribution with $n_F=4$.
The labelling of the various curves is the same as in Fig.~\ref{fig:w1}. We see that after the
resummation of the NLL BFKL corrections the value of $Q_s$ grows much
more slowly than in the LO BFKL case. Moreover, the NLL BFKL curve even drops
below the DGLAP curve, since in the region of $\gamma=0.3\ -\
0.4$ the value of
\begin{equation}
\omega_s^{\rm NLL\ BFKL}<\omega_s^{\rm DGLAP},
\label{eq:ww}
\end{equation}
 see Fig.~\ref{fig:w1}.
For a low value of $\alpha_s$, we find the expected ordering
$$Q_s^{\rm LO ~BFKL} > Q_s^{\rm NLL~BFKL} > Q_s^{\rm DGLAP}$$
in Fig.~\ref{fig:sat}b. However at
larger values of $\alpha_s$, we see $Q_s^{\rm NLL~BFKL}$ becomes less than
$Q_s^{\rm DGLAP}$, as shown in Fig~\ref{fig:sat}c,d.
Due to the fact that the resummed value of $\omega_s$ depends only weakly
on $\alpha_s$,  $\ln Q^2_s$ increases almost linearly with $\ln (1/x)$, with
\begin{equation}
Q^2_s\sim x^{-0.45}.
\label{eq:sx}
\end{equation}
The power of $x$ that we have obtained using NLL resummation is somewhat larger
than the result of \cite{Tr}.
Asymptotically, at very large $Q_s$ (and $\alpha_s(Q^2_s) \ll 1$), the
LO BFKL (dotted) and the resummed NLL BFKL (solid) curves become parallel.

To determine the saturation momentum, $Q_s$, from Fig.~\ref{fig:sat}, we need to
know the value of $x_0$ where, by definition, $Q_s=1$ GeV.  If we were to take
the model of Ref.~\cite{BGBK}, then $x_0$ would be in the
range $0.2 \times 10^{-4} - 0.3 \times 10^{-3}$, see Fig.~8 of \cite{BGBK}.
If we assume that $x_0=10^{-4}$, then we see that the expected saturation
region accessible to the experiments at HERA ($Q^2\sim 2~{\rm GeV}^2,~
x\sim 10^{-5}$) lies outside the pure perturbative domain where
one can safely neglect higher-twist effects, power
corrections, etc.

\section{Conclusions}

The $k_t$ distribution of gluons emitted along the gluon chains
that accompany a hard subprocess at high energy hadron colliders
could give rise to two types of problem.  First on the theoretical
side, we have the possibility of infrared instability at low $k_t$.
Due to the presence of inverse $k_t$-ordering, LO BFKL
evolution tends to populate the low $k_t$ domain where the
QCD coupling $\alpha_s$ is larger.  The other potential problem is
more experimental.   The $k_t$ diffusion, caused by the presence of inverse
$k_t$-ordering, may produce too many large $k_t$ gluons accompanying
the hard subprocess, and, as a consequence, obscure the extraction
of New Physics.

To study these issues, we have used the prescription of Refs.~\cite{Cif2,SPol}
 for NLL BFKL resummation to calculate the resummed
BFKL intercept and the $k_t$ distribution of the intermediate
gluons emitted during the evolution in both the BFKL and DGLAP cases.
We have shown that after such resummation the BFKL intercept becomes
 much smaller than that obtained using the LO BFKL formalism,
and rather close to that obtained using DGLAP evolution.
Indeed, the NLL resummation tames both the potential problems.  The infrared
convergency of the distributions is similar to that for DGLAP.  Secondly, for high $k_t$,
the NLL BFKL gluon distribution only slightly exceeds the DGLAP prediction.
Thus, contrary to what might be expected from LO BFKL studies,
we do not expect any significant increase in accompanying gluon (and therefore minijet and hadron)
transverse momentum over and above that which is obtained using the parton shower Monte Carlo approach.
It turns out that the presence of the terms with inverse ordering in $k_t$ are approximately
compensated by high-twist gluon Reggeization.

Moreover, it follows that the saturation
momentum calculated using the resummed NLL BFKL framework is much lower
than that predicted by LO BFKL. This implies that we
cannot trust the numerical results obtained in the framework of the Balitsky-Kovchegov
equation, which is based on LO BFKL. Therefore the growth of the saturation scale $Q_s$
should in practice be much slower than that predicted, for example, in Refs.~\cite{Levin1,Levin,BNL}.
Indeed from the NLL BFKL curves in Fig.~\ref{fig:sat} we conclude that saturation effects
are not expected in the pure perturbation region accessible to the experiments at HERA.
Even at the LHC, it will not be easy to observe saturation in the perturbative region.
Note that, from (\ref{eq:sx}), we have
\begin{equation}
Q^2_s~\simeq~1 ~{\rm GeV}^2~(x_0/x)^{0.45}~=~1 ~{\rm GeV}^2~(x_0\sqrt s/Q_s)^{0.45}.
\end{equation}
Solving this equation at the LHC energy, $\sqrt{s}=14$ TeV, we obtain in the centre of the pseudorapidity
plateau
\begin{equation}
Q_s(\eta=0)~~\simeq~~1~{\rm GeV}~(x_0\sqrt{s}/1~{\rm GeV})^{0.18} ~~\simeq ~~1.1\ (1.6)~{\rm GeV},
\end{equation}
assuming that $x_0=10^{-4}\ (10^{-3})$. Thus, at the LHC, the only chance to observe
saturation phenomena, with the scale not too low, $Q^2 \sim 10~{\rm GeV}^2$ say,
is to study processes in the fragmentation region, with $x_-\sim 0.1$
and $x_+\sim 10^{-6}$.

The $k_t$ distribution of intermediate gluons emitted along the evolution chain
depends strongly on the form of the input conditions. For a realistic
`Pomeron-like' input the distributions predicted by the resummed NLL
 BFKL and DGLAP evolutions are rather close to each other. However,
because of the contribution of configurations with inverse $k_t$
ordering the NLL BFKL distribution is slightly wider (i.e. has a larger
dispersion).

\section*{Acknowledgements}

We thank Gavin Salam for stimulating this study and for fruitful discussions.
ADM thanks the Leverhulme Trust for an Emeritus Fellowship and MGR thanks the IPPP at the University of
Durham for hospitality. This work was supported by
the UK Particle Physics and Astronomy Research Council, by a Royal Society special
project grant with the FSU, by grant RFBR 04-02-16073
and by the Federal Program of the Russian Ministry of Industry, Science and Technology
SS-1124.2003.2.

\section*{Appendix: equation for the critical anomalous dimension, $\gamma_c$}

The non-linear BFKL evolution equation is\footnote{Here we neglect
the dependence on the momentum transfered through the amplitude
and consider only the component with zero conformal spin $n=0$ (that is,
we consider the forward amplitude
with a flat azimuthal dependence). Recall that the eigenfunctions which have
$n\neq 0$ correspond to $\omega_s<0$~\cite{BFKL} and
therefore the non-linear contribution with $n\neq 0$ will in any case
be small, being suppressed by at least a factor $\alpha_s$, in comparison
with the linear one.}
 \begin{equation}
\frac{df_g(x,k_t)}{dY}=\int dk'^2_t
K(k_t,k'_t)f_g(x,k'_t)- \int dk'^2_t ~V\cdot f^2_g(x,k'_t)
\label{eq:A1}
 \end{equation}
where $K(k_t,k'_t)$
is the BFKL kernel and $V$ is the triple Pomeron vertex.
If we expand
the amplitude in terms of the conformal eigenfunctions
$E_\gamma=e^{\gamma r}$ (for the case of our forward amplitude with
$n=0$),
\begin{equation} f_g=\int e^{\gamma r} F(\gamma,Y)d\gamma,
\label{eq:A2}
\end{equation}
then we obtain the following equation for $F(\gamma,Y)$
\begin{equation}
\frac{dF(\gamma,Y)}{dY}=\omega_s(\gamma)F(\gamma,Y)~-~\int \int v\cdot
F(\gamma_1,Y)F(\gamma_2,Y)d\gamma_1d\gamma_2,
\label{eq:A3}
\end{equation}
where $Y={\rm ln}(1/x)$, and where the triple pomeron vertex in the $\gamma$ representation
is denoted by $v$.
In addition to the extra $\alpha_s$ (in comparison with $\omega_s$), the
vertex $V$ contains the factor $1/R^2Q^2=1/R^2k^2_t$, which represents the small
probability of recombination of two gluons of size
$\sim 1/k_t$ homogeneously distributed over a domain of transverse size $\sim R$.
 When going from Eq.~(\ref{eq:A1}) to Eq.~(\ref{eq:A3}), we have integrated
the product of the vertex $V$ and the two eigenfunctions $E_{\gamma_1}$
and $E_{\gamma_2}$ (hidden in $f_g$) over the transverse momentum $k'_t$.
The integral over $dk'_t$ gives a pole $1/(1+\gamma-\gamma_1 -\gamma_2)$
which reflects the conservation law for the anomalous dimensions.
One of the integrals, say $d\gamma_1$, may be done by
closing the contour on this pole, while the other ($d\gamma_2$) can be
performed by the saddle point method. Because of the symmetry  of the expression, 
the saddle point gives $\gamma_1=\gamma_2$.
 Thus Eq.~(\ref{eq:A3})  takes the form
\begin{equation}
\frac{dF(k,Y)}{dY}=\omega_s(k)F(k,Y)-\bar v F^2(k/2,Y),
\label{eq:A4}
\end{equation}
 where the variables $\gamma$ and $\gamma_i$ are related to the `wave vector'\cite{GLR} $k$ by
\begin{equation}
\gamma ~=~ 1 - k
\label{eq:A3a}
\end{equation}
\begin{equation}
\gamma_1 ~=~ \gamma_2~ = ~\frac{1+\gamma}{2}~=~ 1 - \frac{k}{2}.
\label{eq:A3b}
\end{equation}
 In Eq.~(\ref{eq:A4}), $\bar v$ now denotes the triple Pomeron vertex
together with the measure of integration around the saddle point.

To solve (\ref{eq:A4}) we use the method of characteristics. In the
stationary case, $dF/dY=0$, we have
\begin{equation}
\omega_s(k)F(k,Y)=\bar vF^2(k/2,Y) .
\label{eq:A5}
\end{equation}
The solution of (\ref{eq:A5}) may be written as $F=a(k,Y)e^{fkY}$ where
the factor $f$ in the power of the exponent can be found by matching
the function $F(k,Y)$ in the region of non-linear and linear
evolution, neglecting the  non-linear term in the latter. In the
linear case $fk=\omega_s(k)$. Matching the first derivative\footnote{
Second and higher derivatives can be matched by an appropriate
choice of the pre-exponent factor $a(k,Y)$.},
  we obtain Eq.~(\ref{eq:sat}),
\begin{equation}
d(fk)/dk=f=\omega_s/k=d\omega_s(k)/dk,
\label{eq:A6}
\end{equation}
where, recall, $k=1-\gamma$.

\end{document}